\documentclass[lettersize,journal]{IEEEtran}
\usepackage{amsmath,amsfonts}
\usepackage{algorithmic}
\usepackage{algorithm}
\usepackage{array}
\usepackage[caption=false,font=normalsize,labelfont=sf,textfont=sf]{subfig}
\usepackage{textcomp}
\usepackage{stfloats}
\usepackage{url}
\usepackage{verbatim}
\usepackage{graphicx}
\usepackage{cite}
\begin{document}

\title{Initial On-Sky Performance testing of the Single-Photon Imager for Nanosecond Astrophysics (SPINA) system}

\author{\IEEEauthorblockN{Albert Wai Kit Lau\IEEEauthorrefmark{1},
Nurzhan Shaimoldin\IEEEauthorrefmark{2},
Zhanat Maksut\IEEEauthorrefmark{2}, 
Yan Yan Chan\IEEEauthorrefmark{1}\IEEEauthorrefmark{3},\\
Mehdi Shafiee\IEEEauthorrefmark{2}\IEEEauthorrefmark{4},
Bruce Grossan\IEEEauthorrefmark{2}\IEEEauthorrefmark{5}, and
George F. Smoot\IEEEauthorrefmark{2}\IEEEauthorrefmark{6}\IEEEauthorrefmark{7}\IEEEauthorrefmark{8}\IEEEauthorrefmark{9}\IEEEauthorrefmark{10}\IEEEauthorrefmark{11}}

\IEEEauthorblockA{\IEEEauthorrefmark{1}Department of Physics, The Hong Kong University of Science and Technology, Clear Water Bay, Kowloon, Hong Kong}\\
\IEEEauthorblockA{\IEEEauthorrefmark{2}Energetic Cosmos Laboratory, Nazarbayev University, Astana, Kazakhstan}\\
\IEEEauthorblockA{\IEEEauthorrefmark{3}The William H. Miller III Department of Physics and Astronomy, Johns Hopkins University, Baltimore, MD 21218, USA}\\
\IEEEauthorblockA{\IEEEauthorrefmark{4}Department of Electrical and Computer Engineering, Nazarbayev University, Astana, Kazakhstan}\\
\IEEEauthorblockA{\IEEEauthorrefmark{5}Space Sciences Laboratory, University of California, Berkeley, California, USA}\\
\IEEEauthorblockA{\IEEEauthorrefmark{6}Department of Physics, University of California, Berkeley, California, USA}\\
\IEEEauthorblockA{\IEEEauthorrefmark{7}Institute for Advanced Study, The Hong Kong University of Science and Technology, Clear Water Bay, Kowloon, Hong Kong}\\
\IEEEauthorblockA{\IEEEauthorrefmark{8}Laboratoire APC-PCCP, Université Sorbonne Paris Cité, Université Paris Diderot}\\
\IEEEauthorblockA{\IEEEauthorrefmark{9}Department of Physics, University of California, Berkeley, California, USA}\\
\IEEEauthorblockA{\IEEEauthorrefmark{10}Laboratoire Astroparticule et Cosmologie, Universit{\'e} de Paris, F-75013, Paris, France}\\
\IEEEauthorblockA{\IEEEauthorrefmark{11}Donostia International Physics Center, University of the Basque Country UPV/EHU, E-48080 San Sebastian, Spain}\\

\thanks{Corresponding author: Albert Wai Kit Lau (email: E-mail: awklau@connect.ust.hk}}
% <-this % stops an unwanted space

%\thanks{This paper was produced by the IEEE Publication Technology Group. They are in Piscataway, NJ.}% <-this % stops a space
%\thanks{Manuscript received April 19, 2021; revised August 16, 2021.}}

%{A.W.K. Lau \MakeLowercase{\textit{et al.}}: Initial On-Sky Performance testing of the Single-Photon Imager for Nanosecond Astrophysics (SPINA) system}


\maketitle

\begin{abstract}
This work presents an initial on-sky performance measurement of the Single-Photon Imager for Nanosecond Astrophysics (SPINA) system, part of our Ultra-Fast Astronomy (UFA) program. We developed the SPINA system based on the position-sensitive silicon photomultiplier (PS-SiPM) detector to 
%capture photons with temporal, spatial, and photoelectron (P.E.) counts information. %
record both photoelectron (P.E.) temporal and spatial information. The initial on-sky testing of the SPINA system was successfully performed on UT 2022 Jul 10, on the 0.7-meter aperture Nazarbayev University Transient Telescope at the Assy-Turgen Astrophysical Observatory (NUTTelA-TAO). We measured stars with a wide range of brightness and a dark region of the sky without stars $< 18$ mag. We measured the SPINA system's spatial resolution to be $<232\mu m$ (full-width half-maximum, FWHM), limited by the unstable atmosphere. We measured the total background noise (detector dark counts and sky background) of 1914 counts per second (cps) within this resolution element. We also performed a crosstalk mapping of the detector, obtaining the crosstalk probability of $\sim0.18$ near the detector's center while reaching $\sim 50\%$ at the edges. We derived a $5\sigma$ sensitivity of $17.45$ Gaia-BP magnitude in a 1s exposure with no atmospheric extinction by comparing the received flux with Gaia-BP band data. For a $10ms$ window and a false alarm rate of once per 100 nights, we derived a transient sensitivity of 14.06 mag. For a $1\mu s$ or faster time scale, we are limited by crosstalk to a 15 P.E. detection threshold. In addition, we demonstrated that the SPINA system is capable of capturing changes in the stellar profile FWHM of $\pm1.8\%$ and $\pm5\%$ change in the stellar profile FWHM in $20ms$ and $2ms$ exposures, respectively, as well as capturing stellar light curves on the $ms$ and $\mu s$ scales.
\end{abstract}


\section{Introduction}
In our "Ultra-Fast Astronomy” or UFA program, we aim to survey the night sky in optical bands at timescales rarely explored before, from nanoseconds to milliseconds\cite{li2019program, lau2020sky}. 
%Millisecond transients and variabilities have been detected in various sources, like the Fast Radio Bursts (FRBs) \cite{nunez2021constraining, lau2021constraining}, millisecond pulsars\cite{ambrosino2017optical}, and X-ray binaries \cite{wijnands1998millisecond}.%
Millisecond transient and variable signals  have been detected from Fast Radio Bursts (FRBs) \cite{nunez2021constraining, lau2021constraining} in radio bands, from millisecond pulsars\cite{ambrosino2017optical} in radio, X-ray and gamma-ray bands, and X-ray binaries in x-ray and gamma-ray bands\cite{wijnands1998millisecond}.%
An optical signal related to these transients may exist on a similar time scale or even faster \cite{yang2019bright}. However, the speed of existing CCD cameras limits the research in ultra-fast optical transients to the $ms$ scale \cite{dhillon2007ultracam, harding2016chimera}. Ultra-fast detectors with single photon sensitivity and $ns$ time resolution can push the limit of short timescale optical transient detection. In addition, ultra-fast photon detectors can be used for fast wavefront sensing \cite{zappa2006single}, accurate occultation measurements \cite{richichi2014occultations}, and proposed interstellar communication in optical bands \cite{cosens2018panoramic}.

Recently, we constructed the Single-Photon Imager for Nanosecond Astrophysics (SPINA) system, an instrument utilizing a position-sensitive silicon photomultiplier (PS-SiPM) detector \cite{peng2020square}. In this work, we describe the design of the SPINA system, give the observation log of the first on-sky measurements, and describe our data reduction and processing. We then present the calculation of our sensitivity on different time scales. Additionally, we demonstrate the ability of our system to measure brightness and stellar profiles on the $ms$ and $\mu s$ time scales. Finally, we propose further upgrades to the SPINA system.

\section{The Single-Photon Imager for Nanosecond Astrophysics (SPINA) system}
\label{sec:SPINA}
\begin{figure*}[!t]
    \centering
    \includegraphics[width=0.9\textwidth]{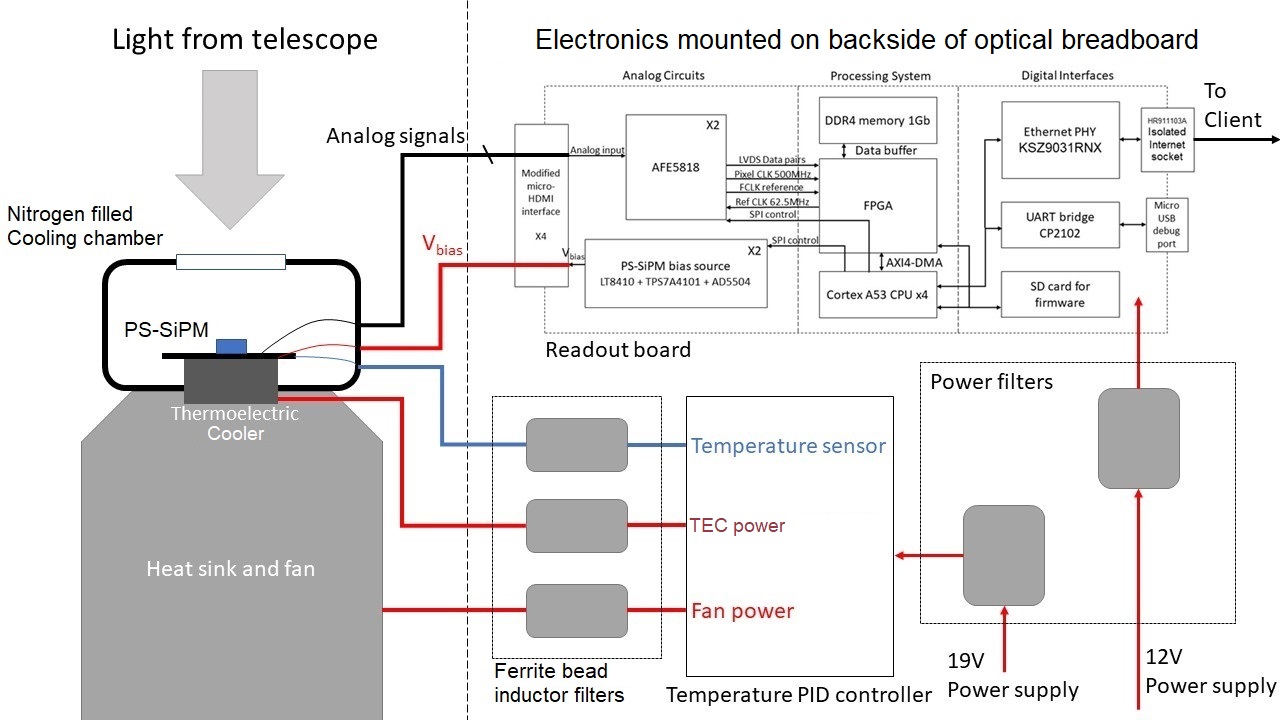}
    \caption{Hardware schematic of the SPINA system}
    \label{fig:schematic}
\end{figure*}

The SPINA electronics were custom built, described in \cite{lau2022development}. A hardware schematic of the SPINA system is shown in Fig. \ref{fig:schematic}. The specifications and operation parameters of the SPINA system are listed in Table \ref{tab:SPINA-Specifications} in Appendix \ref{appendix:spec}.

\subsection{PS-SiPM used in the SPINA system}
We use the PS-SiPM model PSS 11-3030-S from the Novel Device Laboratory in the SPINA system. When a photon is detected by the PS-SiPM detector, a photo-electron (P.E.) is excited and multiplied into a charge pulse. We refer to each charge pulse in our 8 ns time window as a single event. However, when more than one photon converts to a P.E. in a single event, this is called “electron pile-up”.  The signal is greater for more P.E., and so we can use the pulse strength to determine how many P.E. counts were in an event. Therefore, the number of events indicates how many charge pulses are received, and ideally, the number of counts in each event indicates the number of detected photons.

%By comparing pulse strengths on the four anodes, the SPINA system estimates the position of the events and outputs a time series of positions and pulse strength with 8ns time-stamping accuracy. 
Each charge pulse is then collected by the four anodes, one along each of the four sides of the PS-SiPM  \cite{peng2020square}. By comparing pulse strengths on the four anodes, the SPINA system estimates the position of each event and records this position in a 7-bit number each for x and y, and an associated time in 8ns steps. These data are transmitted to the data acquisition computer through an Ethernet interface.

We cool down the detector to reduce dark current using a thermoelectric cooler. We calibrated the dark events rate ($D$) versus the temperature curve prior to the on-sky testing and found: 
\begin{equation}\label{eq:cooling_D}D = 3396\times e^{0.0616T}\end{equation}

\subsection{Mechanical Design}
\label{subsec: Mech}
We needed a mechanical framework for our components that would mount on an instrument port of the Nazarbayev University Transient Telescope at the Assy-Turgen Astrophysical Observatory (NUTTelA-TAO) \cite{grossan2022performance}. We connected a mounting ring to a small optical breadboard that would serve as a mounting platform for our components. The PS-SiPM detector chip and its cooling system are mounted on top of the platform. All control electronics are mounted on the bottom of the platform, which acts as a grounded electromagnetic shield. 

\begin{figure}[!t]
    \centering
    \includegraphics[width=\columnwidth]{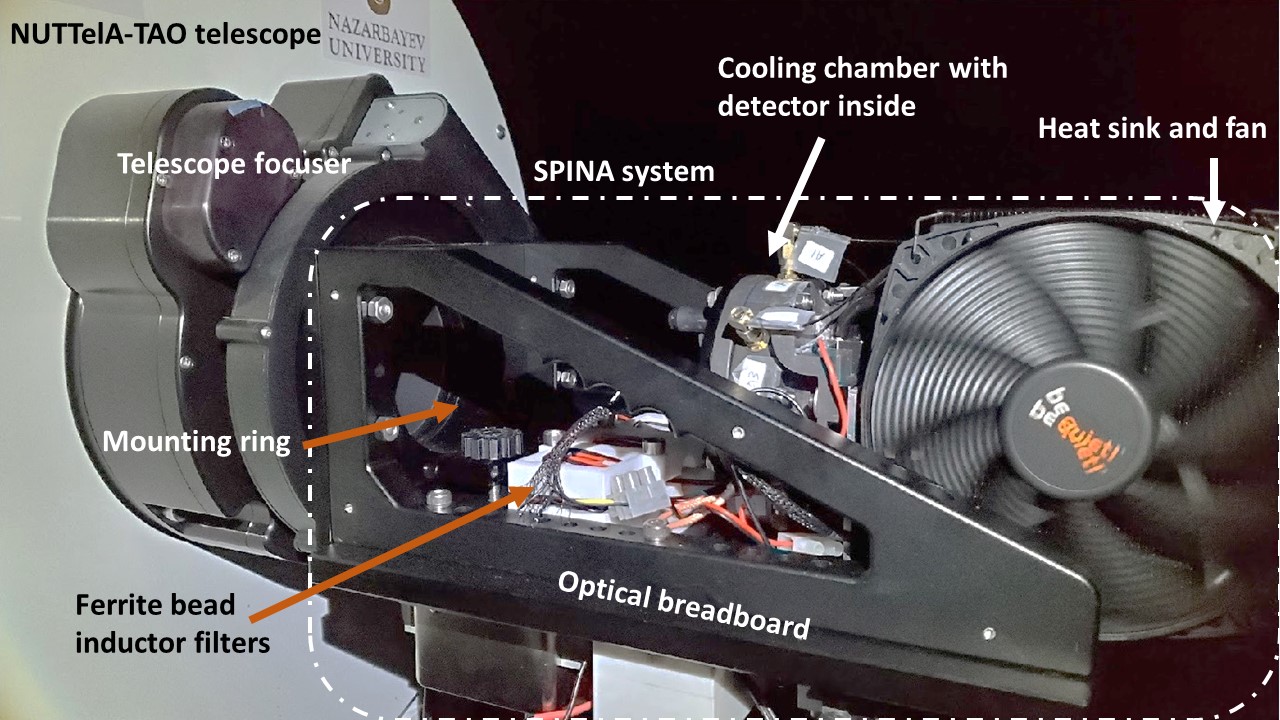}
    \caption{A photograph of the SPINA system with the cover off, shows the configuration of the main components described below as well as electronics and other components in the system schematic in Fig. 1}
    \label{fig:SPINA_internal}
\end{figure}

An optical shield made with painted aluminum sheets covers the testing platform to isolate the SPINA system from stray light. 

\begin{figure}[!t]
    \centering
    \includegraphics[width=0.9\columnwidth]{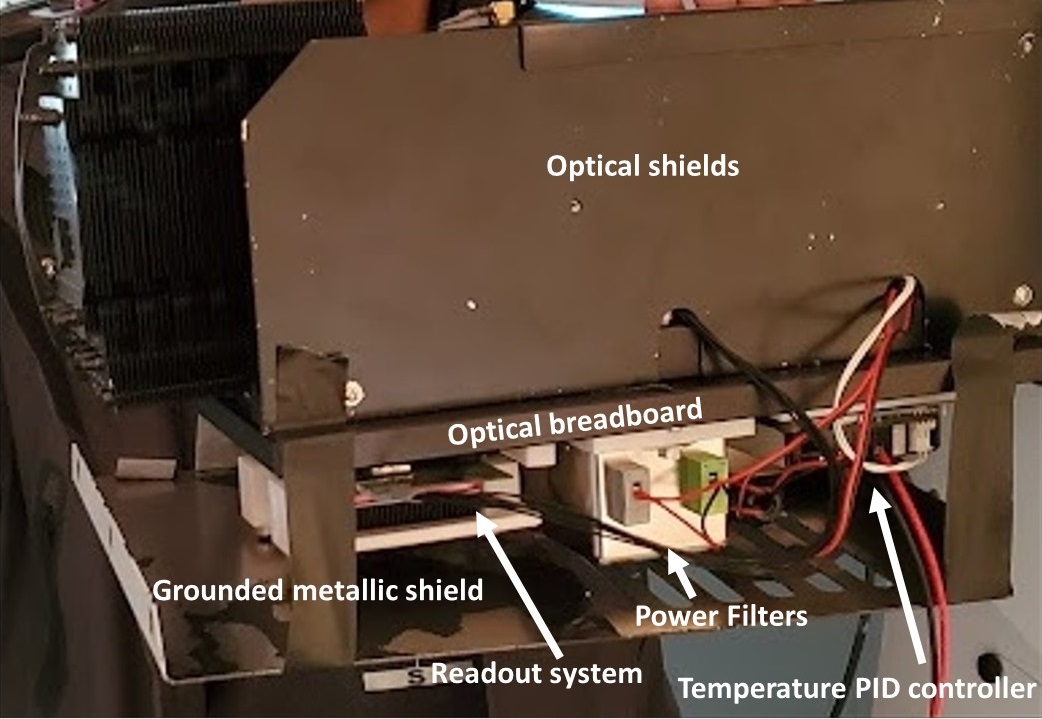}
    \caption{View of the “rear” of the SPINA, the side farthest from the telescope. The figure shows electronic components mounted below the optical breadboard, behind the grounded metal shield.  Above the breadboard, the optical shield is shown (removed in Fig 2).}
    \label{fig:SPINA_shield}
\end{figure}

While we were doing our experiments at the observatory, we found substantial electromagnetic interference (EMI) noise when the telescope mount motors were switched on. To reduce the interference, we built a grounded metallic shield to protect the readout system from EMI, shown in Fig.\ref{fig:SPINA_shield}. 

\section{Observation Details}
\label{sec:obs}
\subsection{Telescope specifications}
\label{subsec:telescope}
We performed the first On-Sky test of the SPINA system on the NUTTelA-TAO telescope. The specifications of the telescope are given in Table \ref{tab:tel_spec} in Appendix \ref{appendix:spec}. The SPINA PS-SiPM has a 2.2716'$\times$2.2716' field of view on this telescope. 

\subsection{2022 July 10 observation}
\label{subsec:July11}
We performed observations during the local nighttime of UT Jul 10. There was a strong thunderstorm at the Assy-Turgen Observatory from UT Jul 10 15:00 to 17:00. The sky became partly clear with some high-level clouds and haze after UT Jul 10 17:30. Our observations began at UT Jul 10 18:56, and last till UT Jul 10 19:55. During observation, the humidity was $90\%$ near the ground, the ambient temperature was $10^{\circ}$C, and the radiometric sky temperature (measured with a Boltwood Cloud detector II) was $-25^{\circ}C\pm5^{\circ}C$, indicating a partly cloudy or hazy sky. The astronomical twilight begins at UT Jul 10 21:07, which did not interfere with our observation. There were also strong wind gusts up to $7ms^{-1}$. Small sections of the sunset sky can make a relatively uniform source of illumination, and can therefore be used to correct non-uniform response of detectors, a "flat field" correction.  Since the sky was cloudy and therefore non-uniform, we could not make the observations for such corrections.

The moon phase was 93.5\%, leading to a bright sky background. The moon was near the Southern horizon during the observation time. Therefore, we pointed the telescope to the Northern sky near the zenith to minimize the effect of the moon, air mass, and the milky way. 

We first focused the system on a relatively bright star just below saturation. We then measured a nearby group of stars near the constellation Draco with a wide range of apparent magnitudes. All stars were within a 5-degree area, and all measurements were performed within 1 hour to minimize atmospheric effects. We chose a sky area without a star brighter than magnitude 18 (Empty Field) to measure the sky background and performed the dark reference measurement with the telescope shutter closed. The observation log is presented in Table \ref{tab:obs_log_28}. We acquired at least 16 million recorded events in each dataset to ensure an adequate signal-to-noise ratio (SNR). 
\begin{table*}
\centering
\caption{Observation log for 2022 July 10}
\label{tab:obs_log_28}
\begin{tabular}{|c|c||c|c|c|}
\hline
Target & Coordinate of pointing & Brightest star mag & Observation Time & Airmass\\
&(J2000 RA, DEC)&(Gaia BP-band)&(Universal Time, UT)&\\
\hline
Focusing 1&19h 12m 37.28s, 67$^{\circ}$ 39' 43.4'' & +3.60 & 18:56 - 19:01 & 1.13\\
Focusing 2&19h 20m 43.43s, 65$^{\circ}$ 42' 53'' & +4.58 & 19:02 - 19:04 & 1.13\\
Star Field 1&19h 12m 37.28s, 67$^{\circ}$ 39' 43.4'' & +3.60 & 19:06 & 1.13\\
Star Field 2&19h 20m 43.43s, 65$^{\circ}$ 42' 53'' & +4.58 & 19:08 & 1.12\\
Star Field 3&18h 56m 28.82s, 65$^{\circ}$ 15' 30.5'' & +5.86 & 19:13 & 1.10\\
Star Field 4&19h 9m 49.13s, 65$^{\circ}$ 58' 44.1'' & +6.25 & 19:16 & 1.11\\
Star Field 5&19h 54m 52.43s, 64$^{\circ}$ 43' 13.6'' & +7.16 & 19:19 & 1.13\\
Star Field 6&19h 51m 16.73s, 65$^{\circ}$ 32' 45.8'' & +8.01 & 19:21 & 1.13\\
Star Field 7&20h 2m 1.03s, 65$^{\circ}$ 53' 22.0'' & +8.46 & 19:23 & 1.14\\
Star Field 8&19h 59m 42.31s, 65$^{\circ}$ 12' 46.6'' & +9.49 & 19:26 & 1.14\\
Star Field 9&19h 52m 7.32s, 64$^{\circ}$ 19' 12.1'' & +10.02 & 19:28 & 1.11\\
Star Field 10&19h 43m 13.44s, 64$^{\circ}$ 31' 27.4'' & +10.51 & 19:30 & 1.11\\
Star Field 11&19h 46m 32.21s, 64$^{\circ}$ 29' 10.3'' & +11.46 & 19:32 & 1.11\\
Star Field 12&19h 47m 19.41s, 64$^{\circ}$ 27' 27.8'' & +12.67 & 19:35 & 1.11\\
Star Field 13&19h 44m 7.31s, 64$^{\circ}$ 36' 33.4'' & +13.01 & 19:37 & 1.11\\
Star Field 14&19h 44m 55.96s, 64$^{\circ}$ 32' 15.5'' & +13.32 & 19:40 & 1.11\\
Star Field 15&19h 45m 53.28s, 64$^{\circ}$ 29' 42.4'' & +14.19 & 19:42 & 1.11\\
Star Field 16&19h 43m 42.94s, 64$^{\circ}$ 31' 0.6'' & +14.87 & 19:44 & 1.10\\
Star Field 17&19h 40m 58.48s, 64$^{\circ}$ 34' 15.5'' & +15.89 & 19:46 & 1.10\\
Dark Reference & shutter closed& $\backslash$ & 19:50 & $\backslash$ \\
Empty Field &19h 44m 15.73s, 64$^{\circ}$ 25' 23.5'' & $>$18 & 19:54 & 1.11\\
\hline
\end{tabular}
\end{table*}  
Throughout the observations, the thermoelectric cooling module kept the detector temperature at $-50\pm0.003^{\circ}C$. We made regular visual inspections to make sure that there was no condensation or frost on the detector's surface or the cooling chamber's window. 

\subsection{Focusing}
\label{subsec:Focus}
The detector was located at the focal plane of the telescope for the measurements. We pointed our system to a 3.58 mag star HIP94376, just below saturation ($\sim3$ mag), for focusing. 

\begin{figure*}[!t]
    \centering
    \includegraphics[width=\textwidth]{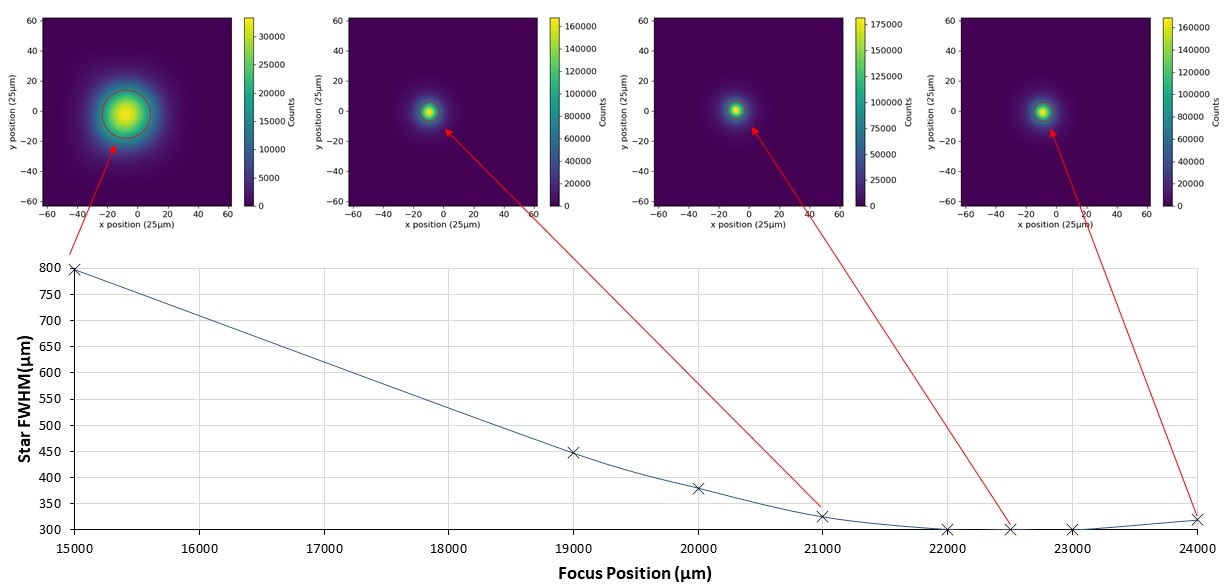}
    \caption{Star FWHM of the SPINA system at different focuser positions}
    \label{fig:foc_plot}
\end{figure*}

The instrument is focused via the telescope focuser, which moves the instrument along the optical path with a travel of 30mm in micron steps. We set an initial position of $15000\mu m$. We captured an image of HIP94376 at multiple positions, fit the star's full-width half-maximum (FWHM) versus the focuser position, as shown in Fig. \ref{fig:foc_plot}. We found the best focus position to be $\sim 22500\mu m$. We frequently checked the focus during the measurements of dimmer stars, and we found no noticeable deviation.  

\section{Data processing}
\begin{figure}[!t]
    \centering
    \includegraphics[width=0.95\columnwidth]{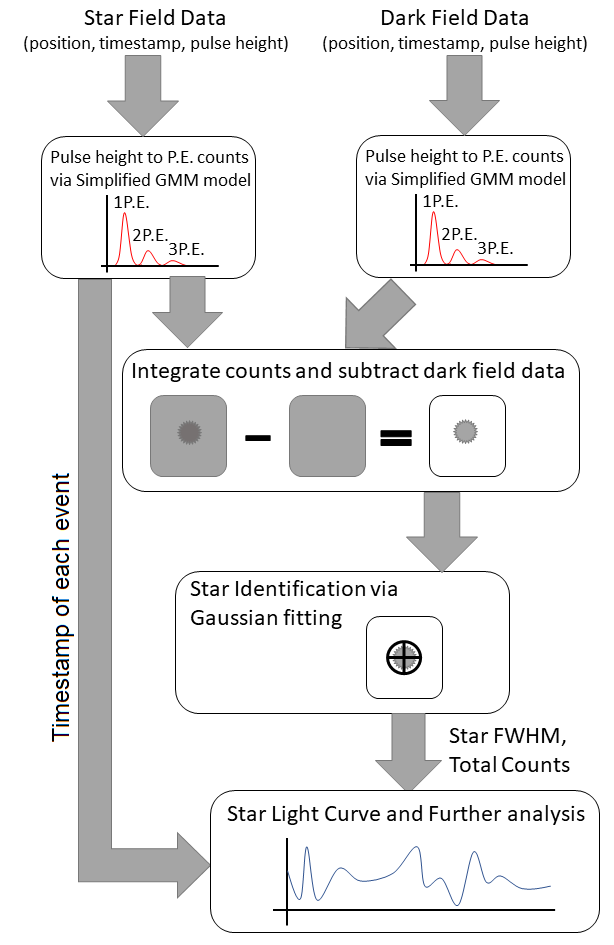}
    \caption{ A schematic of SPINA data processing}
    \label{fig:datapath}
\end{figure}

The SPINA system records each detected photon or noise event with 2-dimensional spatial information (digitized to $25\mu m$ in each dimension), temporal information (timestamped in $8ns$ steps) and pulse strength information (digitized in $15mV$ steps). Note that the spatial sampling is ~ 10X the P.E. localization resolution given in Table \ref{tab:SPINA-Specifications}, and that the fitting results in the next section show that the pulse voltage sampling is also more than adequate. The PS-SiPM does not have separate well-defined detection pixels like some other detectors. However, we will refer to our $25\mu m \times 25 \mu m$ position regions (properly pseudo-pixels) here simply as "pixels", which suffices for the discussion below. A schematic of our data processing is shown in Fig. \ref{fig:datapath}.

\label{sec:processing}
\subsection{Photoelectron identification}
\label{subsec:PE}
In an ideal photon detector, the histogram of output pulse strength should be in discrete steps, as P.E. counts are quantized. In reality, the recorded pulse strength is affected by gain variations across the detector, electronic noise, and imperfect pile-up of photons. Hence the discrete pulse strengths broaden into a mixture of Gaussian distributions. To correctly distinguish the P.E. counts per event, we performed a Gaussian Mixture Model (GMM) fitting on N peaks \cite{alvarez2013design}: 
\begin{equation}P(s) = \sum^N_{n=1} A_n*exp\left[(s-\mu_n)^2/(2\sigma_n^2)\right]\end{equation}
Where $s$ is the pulse strength, $A_n$, $\mu_n$, and $\sigma_n$ are the amplitude, location, and standard deviation of each Gaussian peak.
We reduced the number of dependent parameters of GMM fitting by assuming (1) linear response from the PS-SiPM, (2) standard deviation of each Gaussian peak, $\sigma_n$, growing linearly with the peak number, and (3) higher P.E. events coming from coincidence and crosstalk probability \cite{collazuol2012sipm}. These assumptions are shown in the following equations:

\begin{enumerate}
\item $\mu_n = (\mu_1-\mu_0)*n+\mu_0$, where $\mu_0$ is an offset from readout system electronics. 
\item $\sigma_n = \sigma_1*n$, 
\item $A_n = A_1*C^n$, where $C$ is a constant indicating height ratio of each peak to the previous one. 
\end{enumerate}

By applying these assumptions, we can reduce the number of dependent parameters from $3N$ to $5$ ($\mu_0$, $\mu_1$, $\sigma_1$, $A_1$ and $C$). 

\begin{figure}[!t]
    \centering
    \includegraphics[width=\columnwidth]{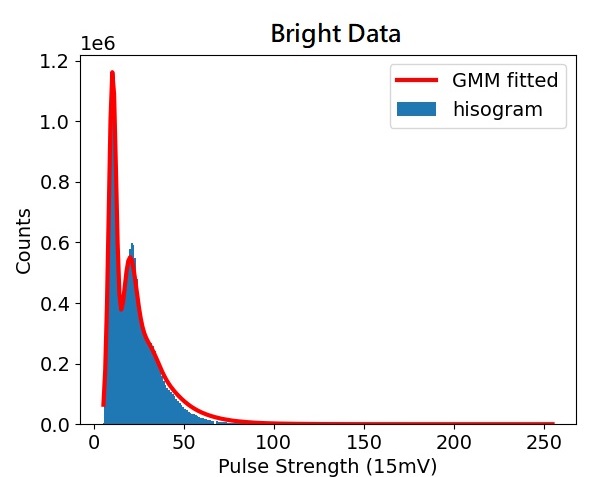}
    \caption{Pulse strength histogram and the GMM fitted result for a 3.6 mag star}
    \label{fig:GMMhist_light}
\end{figure}

\begin{figure}[!t]
    \centering
    \includegraphics[width=\columnwidth]{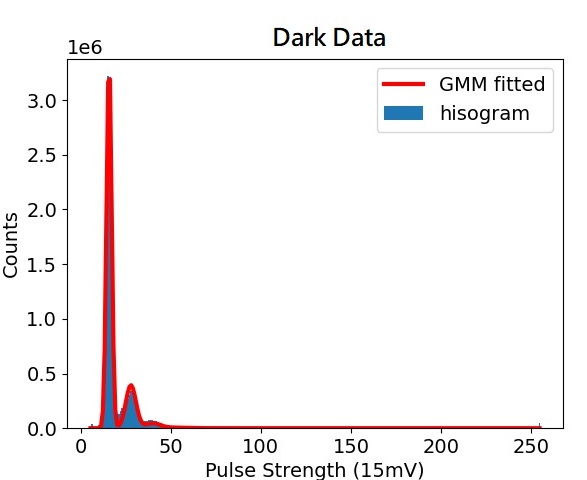}
    \caption{Pulse strength histogram and the GMM fitted result for dark test data (shutter closed)}
    \label{fig:GMMhist_dark}
\end{figure}

The fitting was performed using a non-linear least squares fit algorithm. After fitting the GMM parameters, we matched the strength of each recorded pulse to the most probable number of P.E. counts. Fitting results show that the algorithm works well from dark testing data to the brightest star near saturation, as shown in Fig. \ref{fig:GMMhist_light} and \ref{fig:GMMhist_dark}. 

\subsection{Dark subtraction}
After P.E. identification, the data can be treated as a stream of P.E. counts with spatial information. We reduced the data to 2D images by integrating the P.E. counts according to their pixel position. 

\begin{figure*}[!t]
    \centering
    \includegraphics[width=\textwidth]{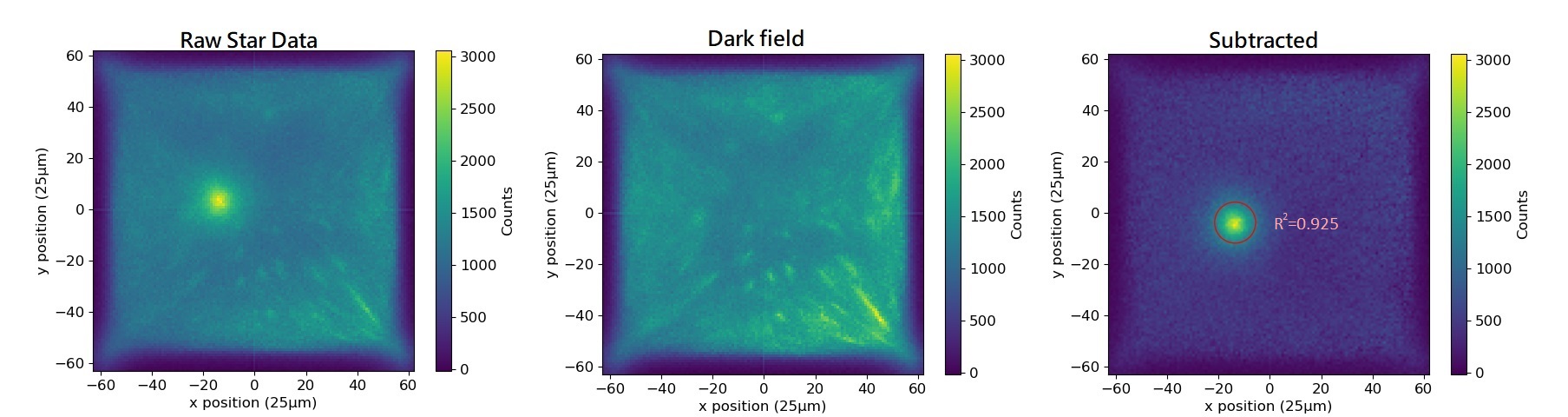}
    \caption{(Left) Integrated image of a 11.46 mag star; (Mid) Integrated image of the empty field; (Right) Subtracted image, with red circle indicating located FWHM of the 11.46 mag star}
    \label{fig:DF_subtraction}
\end{figure*}

The empty field data contain information on detector dark counts and sky background. We subtracted the integrated empty field data from the integrated star field data to obtain a reduced image. The difference in exposure time is compensated with the following equation:
\begin{equation}I_{reduced}(x,y) = I_{star}(x,y) - I_{empty}(x,y)\times \frac{t_{star}}{t_{empty}}\end{equation}
Where $I_{star}$, $I_{empty}$, and $I_{reduced}$ represent integrated images of star field data, empty field data, and subtraction output, respectively. $t_{star}$ and $t_{empty}$ represent the integration time of star and empty field data. A data subtraction sample is shown in Fig. \ref{fig:DF_subtraction}.

\subsection{Star identification}
\label{subsec: star identification}
After calculating the reduced data, we ran a star fitting and identification algorithm. Assuming the star shape is a 2D round Gaussian with an offset (residue sky background from the above subtraction), we obtain the following equation for Gaussian fitting of P.E. counts :
\begin{equation}Count(x,y) = A*exp[((x-\mu_x)^2 + (y-\mu_y)^2))/(2\sigma^2)] + k
\label{Eq: star_fit}
\end{equation}
Where $A$ is the amplitude, $\mu_x$, and $\mu_y$ are the center position of the star, k is the residual background counts, and $\sigma$ is the standard deviation of Gaussian. We performed the fitting with a non-linear least squares fit algorithm. A sample star of 11.46 mag is shown in Fig. \ref{fig:DF_subtraction}, the R-squared value of Gaussian fitting is 0.925, indicating the Gaussian profile fits well.

\section{Sensitivity Analysis}
\label{sec:Sensitivity}

\subsection{Dark Reference and Empty Field frames}
\label{subsec:Dark}
We obtained the Dark Reference data with the telescope shutter closed (Dark Reference in Table \ref{tab:obs_log_28}). We recorded $1.6790161\times10^7$ events within 81.804s, or $205,248\pm50$ events per second. This recorded value is higher than our expectation, as we expected $3396\times e^{-50\times0.0616} \simeq 1.56\times10^{5}$ events per second from Eq. \ref{eq:cooling_D}. 

The excess dark events possibly originated from the scattered moonlight, as the moon was near full (moon phase $93.5\%$), and there were clouds to scatter moonlight. The scattered moonlight may leak into our system through the gaps between photon shield panels. 

We also pointed our system to an area in the sky without a star brighter than 18 mag (Empty Field in Table \ref{tab:obs_log_28}). We recorded $1.6791428\times10^7$ events in 61.043$s$, or $275,116\pm67$ events per second. This background flux comes from the detector noise plus the sky background.

Considering the digitized pixel size of $25\mu m \times 25 \mu m$ from the SPINA system output, we obtained an average overall single-pixel background flux of $19.105$ events per second per pixel. Calculating the P.E. counts per event, we got an average overall single-pixel background flux of 28.898 counts per second ($cps$). 

\subsection{Flux calibration}
\label{subsec:Flux}
We calculated the signal strength of each target star from the fitted Gaussian parameters in Sec. \ref{subsec: star identification}. The calibrated star flux (in cps) can be calculated with the following function:
\begin{equation}Flux = \frac{2\pi A\sigma^2}{t_{bright}}\times 100^{E(AM, AOD)/5}\end{equation}
Where $A$ and $\sigma$ correspond to the amplitude and standard deviation of the fitted Gaussian in Eq. \ref{Eq: star_fit}, $t_{bright}$ corresponds to the exposure time of the star, and $E(AM, AOD)$ corresponds to the extinction calculated with the atmospheric extinction model \cite{green1992magnitude, flanders2008transparency}:
\begin{equation}
\begin{split}
\label{EQ:extinction}
E&(AM, AOD)\\
&= E_{Rayleigh}(AM) + E_{Ozone}(AM) + E_{Aerosol}(AM, AOD)
\\ &= 0.1451AM\times e^{h/8000}\left(\frac{\lambda}{510}\right)^{-4} + 0.016AM \\&+ 2.5AM\times log_{10}\left(e^{AOD\left(\frac{\lambda}{550}\right)^{-1.3}}\right) 
\end{split}
\end{equation}
Where $h = 2750m$ is the altitude of the observatory, $\lambda = 420nm$ is the observation wavelength, which we took to be the PS-SiPM sensitivity peak. AM corresponds to the airmass recorded during observation, and AOD is the aerosol optical depth. Unfortunately, there are no real-time AOD data available, so we can only estimate it: a typical value of $AOD \sim 0.7 - 1$ \cite{flanders2008transparency} for most observatory sites during a hazy night. Following Eq. \ref{EQ:extinction}, we can see the AOD value contributes linearly to the extinction magnitude. All of our observations span a small range of AM ($1.12\pm0.02$). Changing $AOD$ from $1$ to $0.7$ would change the extinction by $0.5$ mag. The effect on linearity, however, is negligible. Therefore, we choose a mean value of $AOD = 0.85$ in the following analysis. 

\subsection{Multi-P.E. probability measurement}
The multi-P.E. events detected by the PS-SiPM either come from the intrinsic crosstalk of the detector itself or the natural coincidence of multiple photons arriving at our detector concurrently. 

\begin{figure}[!t]
    \centering
    \includegraphics[width=\columnwidth]{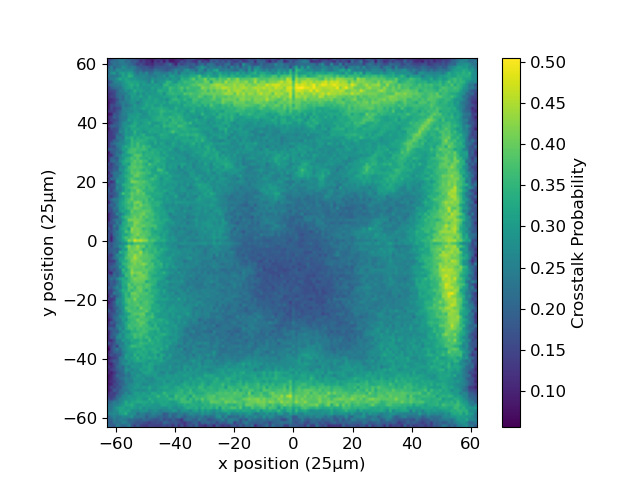}
    \caption{Crosstalk probability map of the PS-SiPM}
    \label{fig:crosstalk_map}
\end{figure}

When intrinsic crosstalk occurs in a PS-SiPM, one detected photon will give a multi-P.E. event. We cannot distinguish this multi-P.E. event from a real multiple photon detection \cite{rosado2015modeling}. 
\begin{equation}
\begin{split}
    &\text{Measured multi-P.E. probability}  = \frac{\text{multi-P.E. event counts}}{\text{total counts}}\\
    & = \text{intrinsic crosstalk probability$+$natural coincidence probability} \\
    & \simeq \text{intrinsic crosstalk probability (low flux condition)}
\end{split}
\end{equation}
We created a crosstalk probability map using the Dark Reference data set, as shown in Fig. \ref{fig:crosstalk_map}. The crosstalk probability is $\sim 0.18$ near the center while reaching $\sim0.5$ near the detector edges. The extra crosstalk near the detector edges possibly comes from photon localization errors near readout electrodes.  

\subsection{Long exposure sensitivity}
We compared our recorded flux to the Gaia DR2 BP-band data, which is sensitive near the blue end of the visible spectrum. 

\begin{figure*}[!t]
    \centering
    \includegraphics[width=0.8\textwidth]{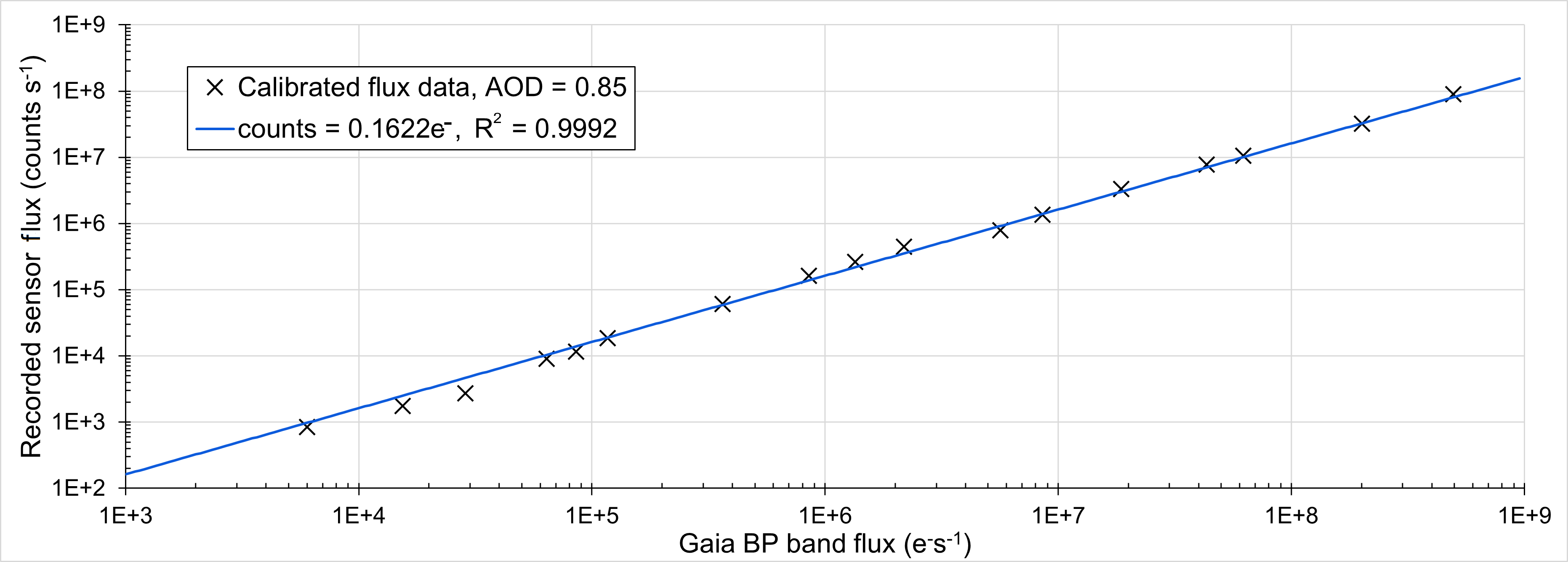}
    \caption{Calibrated detector flux versus Gaia BP-band flux}
    \label{fig:sensitivity}
\end{figure*}

We plotted our calibrated detector flux versus the Gaia BP-band flux to obtain the sensitivity of our system, as shown in Fig. \ref{fig:sensitivity}. We reported the standard flux of Gaia BP-band and our calibrated detector flux in Table \ref{tab:flux_tab} in Appendix \ref{appendix:data}.

In the following sensitivity estimations, we took the best values from the above measurements: $FWHM = 232\mu m$, crosstalk probability= $0.18$ and an overall single-pixel ($25\mu m \times 25\mu m$) background (dark + sky) flux of 19.105 events per second or 28.898 counts per second ($cps$).

We first calculate the FWHM integrated background flux (integrating all background flux within FWHM) $=19.105\pi\left(\frac{232}{2\times25}\right)^2 = 1219$ events per second, or $1914cps$ considering multi-P.E. events. 

Now we can directly calculate the $5\sigma$ detection limit (in $cps$) for long exposures. The signal-to-noise ratio for one second of integration can be calculated with the following equation:
\begin{equation}SNR = \frac{S}{\sqrt{\sigma^2_{dark+sky}+S}}\end{equation}
Solving with $SNR=5$, and using $\sigma^2_{dark+sky}=$ the dark counts and the sky background flux we calculated above, gives a detection limit of $S = 231.063\ cps$, or $1424.56e^-s^{-1}$ Gaia BP-band flux. This corresponds to a limiting magnitude of $17.45$ in the Gaia BP-band (assuming no atmospheric extinction), or $15.68$ mag for $AM=1$ and $AOD=0.85$.

\subsection{Transient sensitivity}
The motivation for developing the SPINA system is the measurement of very short timescale transient events. During the on-sky measurements, as expected, no significant brightness change was observed in the stars. Here, we calculate the detection limits of $10ms$ and $1\mu s$ transients with one false alarm per $5\times10^4 s$ ($\sim 1$ night) and one false alarm per $5\times10^6 s$ ($\sim 100$ night). 

For a $10ms$ bin and a desired false alarm rate of 1 false alarm per $\sim 100$ night, we can accept less than one false alarm per $5\times 10^{9}$ bins, or a false alarm probability of $< 2\times 10^{-10}$ per bin. From Poisson statistics and the 19.14 background counts in one FWHM aperture, this false alarm probability gives a detection limit of 52 counts per 10 ms. From the calibration in Fig. \ref{fig:sensitivity}, the transient should give at least $320.59e^-$ in Gaia BP-band in 10ms, corresponding to a 14.06 mag burst for 10ms. If we loosen the accepted false alarm rate to one false alarm per night (false alarm probability of $< 2\times 10^{-8}$ per bin), we will need a detection limit of $47$ counts per 10ms bin, corresponding to a 14.18 mag burst within 10ms in Gaia BP-band.  

For fast transient events within a $1\mu s$ bin, we consider the FWHM integrated background flux of $1219$ events per second (we used events instead of counts here to avoid double counting the crosstalk effect). We expect an average of only one event per $\sim 800$ $1\mu s$ bins. However, the crosstalk effect will limit our detection potential. We recorded a minimum crosstalk probability of $\sim 0.18$ near the center of the detector, which is much higher than the possibility of having a count in the time bin. Therefore, we need to calculate the crosstalk effect with the geometric statistic \cite{rosado2015modeling}. For a desired false alarm rate of 1 false alarm per $\sim 100$ nights, we can accept less than one false alarm per $5\times 10^{13}$ bins, or a false alarm probability of $2\times 10^{-14}$ per bin. We can then calculate: $1.219\times10^{-3}\times (0.18)^N \leq 2\times 10^{-14}$, where $N$ is the minimum counts detected in a bin for triggering. Solving gives a trigger threshold of at least 15 counts detected within a $1\mu s$ bin. If we loosen the accepted false alarm rate to one false alarm per night (false alarm probability of $< 2\times 10^{-12}$ per bin), we will need a detection limit of 12 counts detected within a $1\mu s$ bin. 

\section{Atmospheric Effects}
\label{sec:Atm_effect}
\begin{figure}[!t]
    \centering
    \includegraphics[width=\columnwidth]{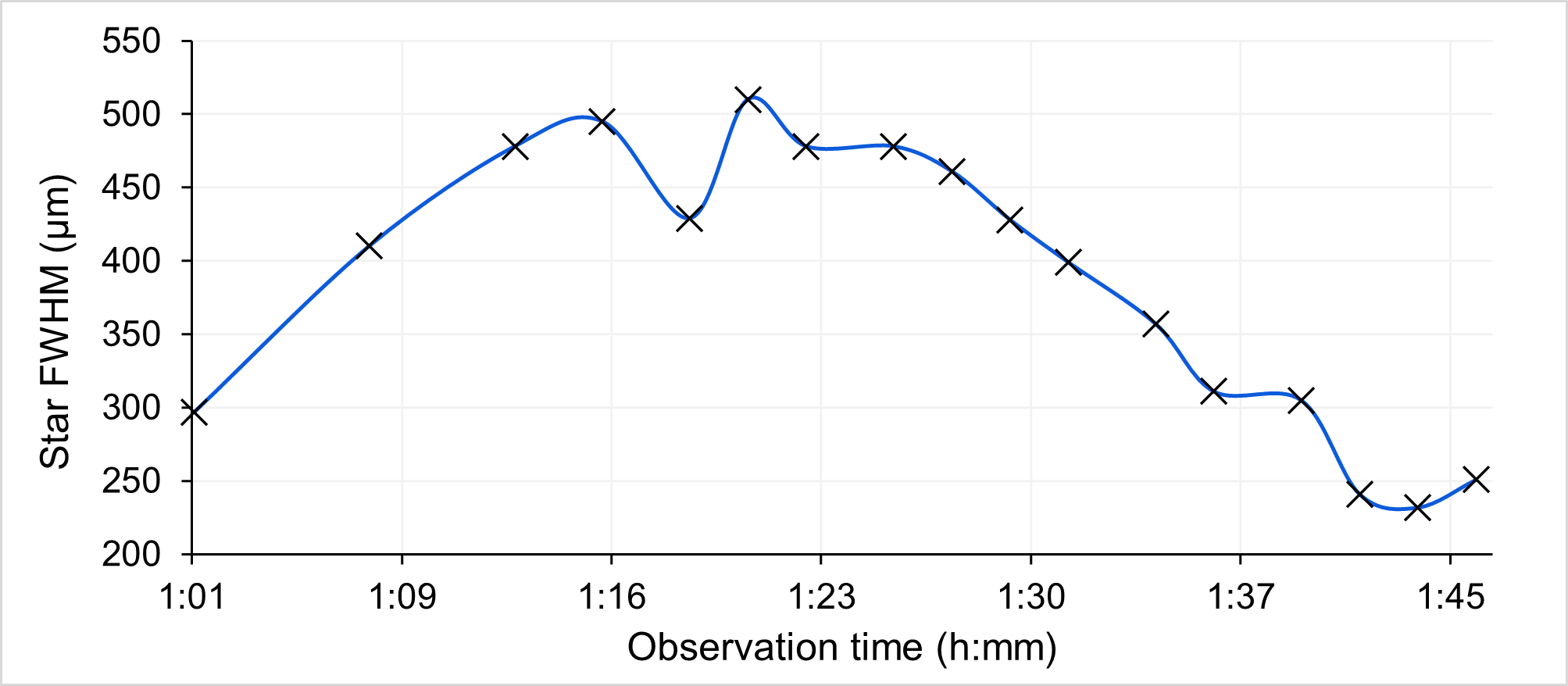}
    \caption{Observed Star FWHM by the SPINA system throughout the observation session}
    \label{fig:FWHM_time}
\end{figure}

Throughout observation, we found that due to atmospheric instability, the star profiles vary in different data sets. We obtained a best FWHM of $232\mu m$ near the end of the observation session and a worst FWHM of $510\mu m$ mid-session. The FWHM versus time is plotted in Fig \ref{fig:FWHM_time}.

\begin{figure}[!t]
    \centering
    \includegraphics[width=\columnwidth]{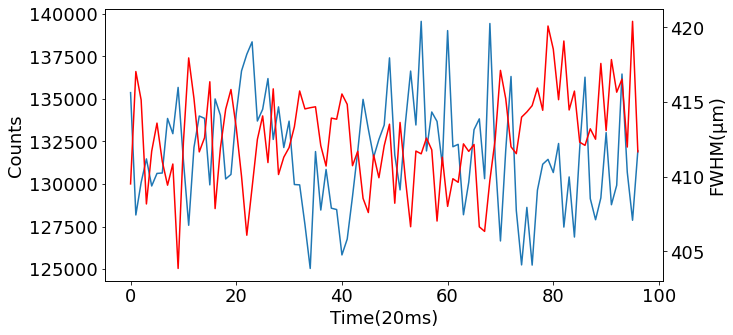}
    \caption{$20ms$ binned FWHM (red) and light curve (blue) of a 4.58 mag star}
    \label{fig:4p6_20ms_FWHM}
\end{figure}

\begin{figure}[!t]
    \centering
    \includegraphics[width=\columnwidth]{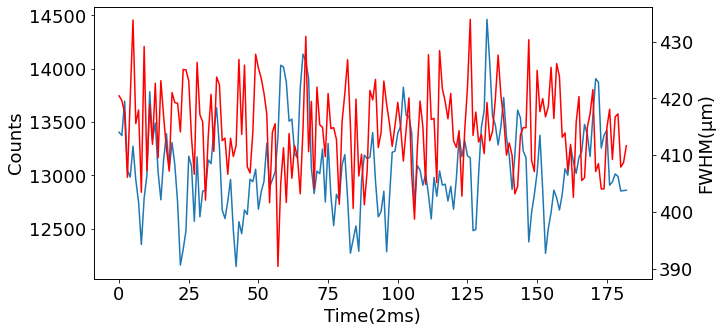}
    \caption{$2ms$ binned FWHM (red) and light curve (blue) of a 4.58 mag star}
    \label{fig:4p6_2ms_FWHM}
\end{figure}

The SPINA system allows us to investigate atmospheric instability with a much faster timescale. We put the recorded data of the 4.58 mag star HIP95081 (Star Field 2) into $20ms$ and $2ms$ bins, and analyzed the FWHM and photon counts per bin following the data processing method in Sec. \ref{sec:processing} (The photon counts per time bin are calculated with the FWHM
of the corresponding time bin for aperture correction.). We plotted the FWHM and photon counts per bin in Figs. \ref{fig:4p6_20ms_FWHM} and \ref{fig:4p6_2ms_FWHM}. In the $20ms$ binned plot, we measured a maximum $\pm 5.6\%$ fluctuation in photon counts and $\pm 1.8\%$ fluctuation in FWHM. For the $2ms$ binned plot, we measured maximum $\pm 9.4\%$ fluctuation in photon counts and $\pm 5\%$ fluctuation in FWHM. 

The observed fluctuations surpass the anticipated Poisson noise of photons. The expected standard deviation from Poisson noise should amount to approximately 0.275\%  ($\sqrt{132500} \sim 364$ counts) in the $20ms$ binned light curve and around 0.868\% ($\sqrt{13250} \sim 115$ counts) in the $2ms$ binned light curve. The recorded fluctuations are an order of magnitude larger than the expected Poisson noise, and therefore must be due to some other effect.

\begin{figure}[!t]
    \centering
    \includegraphics[width=0.8\columnwidth]{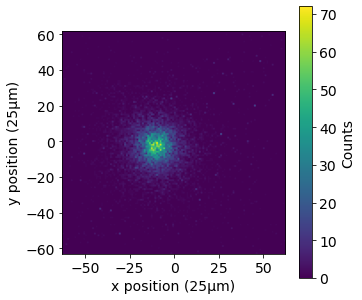}
    \caption{A 4.58 mag star profile from $2ms$ of data. Compare to Fig. 8, right}
    \label{fig:4p6_2ms_img}
\end{figure}

We can see a rough anti-correlation between the FWHM and photon counts: generally, the FWHM peaks when the recorded counts dip. This anti-correlation might be caused by scattering by clouds and haze, which reduces the photon counts from the target star, while broadening the star profile. A sample star profile in a $2ms$ time bin is shown in Fig. \ref{fig:4p6_2ms_img}. 

These data show the SPINA system's capability of capturing the fast-changing light curve and profile of a bright star on $\sim ms$ time scale, comparable to the state-of-the-art electron-multiplying charge-coupled devices (EMCCD) for wavefront sensing \cite{close2012first}. The SPINA system, or photon imagers in general, may act as wavefront detectors when coupled with suitable optics. 

\section{High time-resolution light curve of stars}
\label{sec:time_res_lightcurve}
\begin{figure*}[!t]
    \centering
    \includegraphics[width=0.8\textwidth]{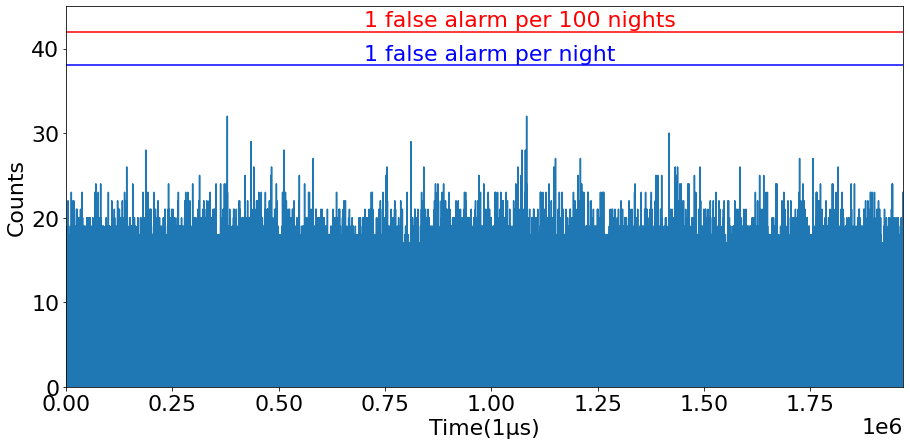}
    \caption{$1\mu s$ binned light curve of a 4.58 mag star}
    \label{fig:4p6_1us_img}
\end{figure*}

One of the a priori goals of the SPINA system was to measure high-time resolution light curves. We plot a $1\mu s$ binned light curve of the 4.58 mag star HIP95081 (Star Field 2) in Fig. \ref{fig:4p6_1us_img} with the propagated detection limits for one false alarm per night and one false alarm per 100 nights. We recorded no significant changes in brightness from HIP95081, as expected. The ability to measure high time-resolution light curves of stars demonstrated that the SPINA system can capture transients from celestial objects and perform accurate occultation measurements.

\section{Future Improvements of the SPINA system}
\label{sec: Future}
For a future long-term observation campaign, we propose the following improvements to the SPINA system.

\subsection{Lowering detector operation temperature}
\label{subsec: Cooling}
Even at the current $-50^\circ C$ detector temperature, there is still a significant dark count rate from the PS-SiPM detector. We would therefore like to increase the cooling system capacity. We plan to install a $100W$ thermoelectric cooler and aim to get the detector temperature to $-70^\circ C$. We expect a roughly 4-fold improvement of dark counts to $<50kcps$. Further cooling may also help reduce the crosstalk probability. 

\subsection{Improving optical and electromagnetic shielding of detector}
\label{subsec: future_shielding}
We found more dark counts in the Dark Reference measurement than expected. We suspect the extra dark counts came from scattered moonlight, which might leak into the system through slits and feed-through holes between optical shields. To avoid this, we are designing an improved optical shield to be made with a single piece of aluminum tube, directly connecting the telescope flange to the detector chamber. Multiple baffles can also be added to reduce internal stray light further. 

During this experiment, we built a temporary metallic EMI shield to tackle the intense EMI noise from the telescope motors. In the future, we will enclose our system in a permanent, sealed metal box to ensure the system works well in harsh environments with EMI and moisture. 
\subsection{Laboratory characterization of the SPINA system}
\label{subsec: In-Lab}
To further characterize the performance of the SPINA system, we have built a light injection setup in the laboratory. This consists of a broad-spectrum lamp that provides optical power through a monochromator and an integrating sphere that provides a high neutral density (ND) ratio with real-time optical power monitoring. We plan to calibrate the absolute quantum efficiency from 300 to 1100nm. We will also use a picosecond pulse laser to examine the time-stamping accuracy of the system. 

With the characterization data, we will optimize the operating parameters (bias voltage, filter bands, etc.) of the SPINA system to obtain an optimal balance between P.E. gain, spatial resolution, dark counts and crosstalk. 

\subsection{Multi-timescales Detection algorithm for fast transient signals}
\label{subsec: time-algo}
In Sec.\ref{sec:Sensitivity}, we assumed some timescales (e.g. $10ms$ and $1\mu s$ bins) to calculate threshold levels of detection. Despite lacking prior knowledge regarding the expected timescales of astronomical transients, it is crucial to develop a multi-timescale model to investigate the potential presence of significant transient signals within the measured photon stream. To achieve this, we can employ a binning strategy with increasing time intervals, such as $8ns,\ 16ns,\ 32ns$, and so on, while establishing a trigger threshold for each timescale. Moreover, the SPINA system generates a considerable volume of data (approximately $100 Mbps$, with 64 bits per event, under a typical 1.5 million events per second flux on the detector). Implementing a real-time event differentiation algorithm is essential for selecting potential transient data, thus effectively reducing data storage and internet bandwidth demands on our system.

\section{Conclusion}
The first on-sky measurements with the SPINA system have demonstrated the ability to measure rapid brightness fluctuations of celestial objects down to a time scale of $\mu s$. This provides a proof of concept for using position-sensitive silicon photomultiplier sensors in astronomical observations.

While EMCCDs are limited to the $\sim100 ms$ regime, PS-SiPM detectors can probe below the microsecond regime. Thus far, however, they do not match EMCCDs in terms of quantum efficiency, pixel count, or red response. We expect these parameters to improve. Manufacturers are now particularly working to improve red response, important to both sample this part of the spectrum (e.g. in filter-band defined observations) and because many astrophysical sources are subject to extinction, which is much larger at blue (shorter) wavelengths.

With the successful testing of the SPINA system, we plan to scale up the imaging area by creating mosaics of SiPM detectors to increase the field of view and increase the likelihood of capturing optical transient events. In addition, multiple observation stations may be deployed to eliminate random signals via coincidence. We believe the Ultra-Fast detectors arrays on multiple observation stations will significantly advance our understanding of the sky on short timescales.

\section{Acknowledgements}
\noindent We acknowledge support from the Jockey Club Institute for Advanced Study (IAS), Hong Kong University of Science and Technology, and from the Energetic Cosmos Laboratory, Nazarbayev University.

\noindent This research was partially funded by the Science Committee of the Ministry of Science and Higher Education of the
Republic of Kazakhstan (Grant No. AP14870504).

\noindent We would like to offer special thanks to the staffs of the Assy-Turgen Observatory and the Fesenkov Astrophysical Institute, especially Dr. Maxim Krugov, for their support, help and advice during the experiment.

\noindent We wish to acknowledge help in preparing the experiment by technicians Ulf Lampe and TK Cheng of the HKUST Physics Department.

\noindent This work has made use of data from the European Space Agency (ESA) mission
{\it Gaia} (\url{https://www.cosmos.esa.int/gaia}), processed by the {\it Gaia}
Data Processing and Analysis Consortium (DPAC, \url{https://www.cosmos.esa.int/web/gaia/dpac/consortium}). 
Funding for the DPAC has been provided by national institutions, in particular the institutions participating in the {\it Gaia} Multilateral Agreement.

%\section{References Section}
%You can use a bibliography generated by BibTeX as a .bbl file.
% The IEEEtran BibTeX style support page is:
 

\bibliographystyle{IEEEtran}
\bibliography{ref}

\begin{thebibliography}{10}
\providecommand{\url}[1]{#1}
\csname url@samestyle\endcsname
\providecommand{\newblock}{\relax}
\providecommand{\bibinfo}[2]{#2}
\providecommand{\BIBentrySTDinterwordspacing}{\spaceskip=0pt\relax}
\providecommand{\BIBentryALTinterwordstretchfactor}{4}
\providecommand{\BIBentryALTinterwordspacing}{\spaceskip=\fontdimen2\font plus
\BIBentryALTinterwordstretchfactor\fontdimen3\font minus
  \fontdimen4\font\relax}
\providecommand{\BIBforeignlanguage}[2]{{%
\expandafter\ifx\csname l@#1\endcsname\relax
\typeout{** WARNING: IEEEtran.bst: No hyphenation pattern has been}%
\typeout{** loaded for the language `#1'. Using the pattern for}%
\typeout{** the default language instead.}%
\else
\language=\csname l@#1\endcsname
\fi
#2}}
\providecommand{\BIBdecl}{\relax}
\BIBdecl

\bibitem{li2019program}
S.~Li, G.~F. Smoot, B.~Grossan, A.~W.~K. Lau, M.~Bekbalanova, M.~Shafiee, and
  T.~Stezelberger, ``Program objectives and specifications for the ultra-fast
  astronomy observatory,'' in \emph{AOPC 2019: Space Optics, Telescopes, and
  Instrumentation}, vol. 11341.\hskip 1em plus 0.5em minus 0.4em\relax SPIE,
  2019, pp. 513--521.

\bibitem{lau2020sky}
A.~W. Lau, M.~Shafiee, G.~F. Smoot, B.~Grossan, S.~Li, and Z.~Maksut, ``On-sky
  silicon photomultiplier detector performance measurements for millisecond to
  sub-microsecond optical source variability studies,'' \emph{Journal of
  Astronomical Telescopes, Instruments, and Systems}, vol.~6, no.~4, p. 046002,
  2020.

\bibitem{nunez2021constraining}
C.~N{\'u}{\~n}ez, N.~Tejos, G.~Pignata, C.~D. Kilpatrick, J.~X. Prochaska,
  K.~E. Heintz, K.~Bannister, S.~Bhandari, C.~Day, A.~Deller \emph{et~al.},
  ``Constraining bright optical counterparts of fast radio bursts,''
  \emph{Astronomy \& Astrophysics}, vol. 653, p. A119, 2021.

\bibitem{lau2021constraining}
A.~W.~K. Lau, A.~Mitra, M.~Shafiee, and G.~Smoot, ``Constraining heii
  reionization detection uncertainties via fast radio bursts,'' \emph{New
  Astronomy}, vol.~89, p. 101627, 2021.

\bibitem{ambrosino2017optical}
F.~Ambrosino, A.~Papitto, L.~Stella, F.~Meddi, P.~Cretaro, L.~Burderi,
  T.~Di~Salvo, G.~L. Israel, A.~Ghedina, L.~Di~Fabrizio \emph{et~al.},
  ``Optical pulsations from a transitional millisecond pulsar,'' \emph{Nature
  Astronomy}, vol.~1, no.~12, pp. 854--858, 2017.

\bibitem{wijnands1998millisecond}
R.~Wijnands and M.~Van Der~Klis, ``A millisecond pulsar in an x-ray binary
  system,'' \emph{nature}, vol. 394, no. 6691, pp. 344--346, 1998.

\bibitem{yang2019bright}
Y.-P. Yang, B.~Zhang, and J.-Y. Wei, ``How bright are fast optical bursts
  associated with fast radio bursts?'' \emph{The Astrophysical Journal}, vol.
  878, no.~2, p.~89, 2019.

\bibitem{dhillon2007ultracam}
V.~Dhillon, T.~Marsh, M.~Stevenson, D.~Atkinson, P.~Kerry, P.~Peacocke,
  A.~Vick, S.~Beard, D.~Ives, D.~Lunney \emph{et~al.}, ``Ultracam: an
  ultrafast, triple-beam ccd camera for high-speed astrophysics,''
  \emph{Monthly Notices of the Royal Astronomical Society}, vol. 378, no.~3,
  pp. 825--840, 2007.

\bibitem{harding2016chimera}
L.~K. Harding, G.~Hallinan, J.~Milburn, P.~Gardner, N.~Konidaris, N.~Singh,
  M.~Shao, J.~Sandhu, G.~Kyne, and H.~E. Schlichting, ``Chimera: a wide-field,
  multi-colour, high-speed photometer at the prime focus of the hale
  telescope,'' \emph{Monthly Notices of the Royal Astronomical Society}, vol.
  457, no.~3, pp. 3036--3049, 2016.

\bibitem{zappa2006single}
F.~Zappa, S.~Tisa, S.~Cova, P.~Maccagnani, D.~B. Calia, R.~Saletti,
  R.~Roncella, G.~Bonanno, and M.~Belluso, ``Single-photon avalanche diode
  arrays for fast transients and adaptive optics,'' \emph{IEEE transactions on
  instrumentation and measurement}, vol.~55, no.~1, pp. 365--374, 2006.

\bibitem{richichi2014occultations}
A.~Richichi, O.~Fors, F.~Cusano, and V.~Ivanov, ``Final binary stars results
  from the vlt lunar occultations program,'' \emph{The Astronomical Journal},
  vol. 147, p.~5, 01 2014.

\bibitem{cosens2018panoramic}
M.~Cosens, J.~Maire, S.~A. Wright, F.~Antonio, M.~Aronson, S.~A.
  Chaim-Weismann, F.~D. Drake, P.~Horowitz, A.~W. Howard, R.~Raffanti
  \emph{et~al.}, ``Panoramic optical and near-infrared seti instrument:
  prototype design and testing,'' in \emph{Ground-based and Airborne
  Instrumentation for Astronomy VII}, vol. 10702.\hskip 1em plus 0.5em minus
  0.4em\relax International Society for Optics and Photonics, 2018, p. 107025H.

\bibitem{peng2020square}
Y.~Peng, W.~Lv, L.~Dai, T.~Zhao, K.~Liang, R.~Yang, and D.~Han, ``A
  square-bordered position-sensitive silicon photomultiplier toward
  distortion-free performance with high spatial resolution,'' \emph{IEEE
  Electron Device Letters}, vol.~41, no.~12, pp. 1802--1805, 2020.

\bibitem{lau2022development}
A.~W.~K. Lau, Y.~Y. Chan, M.~Shafiee, G.~F. Smoot, and B.~Grossan,
  ``Development of position-sensitive photon-counting imager for ultra-fast
  astronomy,'' in \emph{X-Ray, Optical, and Infrared Detectors for Astronomy
  X}, vol. 12191.\hskip 1em plus 0.5em minus 0.4em\relax SPIE, 2022, pp.
  312--329.

\bibitem{grossan2022performance}
B.~Grossan, Z.~Maksut, T.~Komesh, and M.~Krugov, ``Performance of the
  nuttela-tao instrument system after two years of operation,'' in
  \emph{Ground-based and Airborne Instrumentation for Astronomy IX}, vol.
  12184.\hskip 1em plus 0.5em minus 0.4em\relax SPIE, 2022, pp. 2678--2685.

\bibitem{alvarez2013design}
V.~{\'A}lvarez, M.~Ball, F.~Borges, S.~C{\'a}rcel, J.~Castel, S.~Cebrian,
  A.~Cervera, C.~Conde, T.~Dafni, T.~Dias \emph{et~al.}, ``Design and
  characterization of the sipm tracking system of next-demo, a demonstrator
  prototype of the next-100 experiment,'' \emph{Journal of Instrumentation},
  vol.~8, no.~05, p. T05002, 2013.

\bibitem{collazuol2012sipm}
G.~Collazuol, ``The sipm physics and technology-a review,'' \emph{LAL Orsay},
  2012.

\bibitem{green1992magnitude}
D.~W. Green, ``Magnitude corrections for atmospheric extinction,''
  \emph{International Comet Quarterly}, vol.~14, p.~55, 1992.

\bibitem{flanders2008transparency}
T.~Flanders and P.~Creed, ``Transparency and atmospheric extinction,'' 2008.

\bibitem{rosado2015modeling}
J.~Rosado, V.~M. Aranda, F.~Blanco, and F.~Arqueros, ``Modeling crosstalk and
  afterpulsing in silicon photomultipliers,'' \emph{Nuclear Instruments and
  Methods in Physics Research Section A: Accelerators, Spectrometers, Detectors
  and Associated Equipment}, vol. 787, pp. 153--156, 2015.

\bibitem{close2012first}
L.~M. Close, J.~R. Males, D.~A. Kopon, V.~Gasho, K.~B. Follette, P.~Hinz,
  K.~Morzinski, A.~Uomoto, T.~Hare, A.~Riccardi \emph{et~al.}, ``First
  closed-loop visible ao test results for the advanced adaptive secondary ao
  system for the magellan telescope: Magao's performance and status,'' in
  \emph{Adaptive Optics Systems III}, vol. 8447.\hskip 1em plus 0.5em minus
  0.4em\relax SPIE, 2012, pp. 357--372.

\end{thebibliography}
%


%If you have an EPS/PDF photo (graphicx package needed), extra braces are needed around the contents of the optional argument to biography to prevent the LaTeX parser from getting confused when it sees the complicated $\backslash${\tt{includegraphics}} command within an optional argument. (You can create your own custom macro containing the $\backslash${\tt{includegraphics}} command to make thing simpler here.)
 

%\bf{If you include a photo:}\vspace{-33pt} 
%\begin{IEEEbiography}[{\includegraphics[width=1in,height=1.25in,clip,keepaspectratio]{fig1}}]{Testing} Use $\backslash${\tt{begin\{IEEEbiography\}}} and then for the 1st argument use $\backslash${\tt{includegraphics}} to declare and link the author photo. Use the author name as the 3rd argument followed by the biography text.
%\end{IEEEbiography}

%\begin{IEEEbiographynophoto}{Testing2}
\newpage
\onecolumn

\appendices
\section{Specifications}
\label{appendix:spec}

\begin{table*}[h!]
\centering
\caption{Specifications and operation parameters of the SPINA system} 
\label{tab:SPINA-Specifications}
\begin{tabular}{ll}
\hline
Parameter & Value \\
\hline 
Detector model & PSS 11-3030-S \\
Active area & $3.0mm\times3.0mm$\\
Overvoltage ($V_{OV}$) & $9V$\\
Sensitivity peak & 420nm\\
Peak photon detection efficiency & 32\% \\
Sensitive Band (photon detection efficiency $>$ 10\%) & 350nm-600nm\\ 
Detector temperature & $-50^{\circ}C\pm0.003^{\circ}C$ \\
Expected dark event rate & 156kcps\\
Gain & $3.6\times 10^5$\\
Expected single photon spatial resolution (FWHM) & $\sim 250\mu$m\\
\hline
\end{tabular}
\end{table*}

\begin{table*}[h!]
\centering
\caption{NUTTelA-TAO Specifications}
\label{tab:tel_spec}
\begin{center}  
\begin{tabular}{ll}
\hline
Parameter & Value \\
\hline 
Diameter of primary mirror& 0.7m\\
Central obstruction	& 47\% of primary mirror diameter\\
Effective light collection area & 0.2998m$^2$\\
Focal length & 4540mm (F6.5) \\
 Optimal field of view & 70mm (0.86$^{\circ}$)\\
Image scale	 & 22$\mu m$ per arcsecond\\
Number of Nasmyth ports & 2\\
\hline
\end{tabular}
\end{center}
\end{table*}

\section{Flux data reporting}
\label{appendix:data}
\begin{table*}[h!]
\centering
\caption{Standard flux of Gaia BP-band and calibrated detector flux of all observed stars}
\label{tab:flux_tab}
\begin{tabular}{ccccc}
\hline
Target&Brightest star mag & Reference star flux&Recorded star FWHM& Calibrated detector flux\\
&(Gaia BP-band)&($e^-s^{-1}$)&($\mu m$)&(cps)\\
\hline
Star Field 1 & +3.60 & 494140201 & 297 & 89815654\\
Star Field 2 & +4.58 & 200398312 & 410 & 32168011 \\
Star Field 3 & +5.86 & 62057002 & 478 & 10629780\\
Star Field 4 & +6.25 & 43344361 & 495 & 7703222\\
Star Field 5 & +7.16 & 18611289 & 429 & 3299471\\
Star Field 6 & +8.01 & 8561425 & 510 & 1348934\\
Star Field 7 & +8.46 & 5653797 & 478 & 788271\\
Star Field 8 & +9.49 & 2183904 & 478 & 442759\\
Star Field 9 & +10.02 & 1346967 & 461 & 261786\\
Star Field 10 & +10.51 & 850558 & 428 & 162248\\
Star Field 11 & +11.46 & 363960 & 399 & 60745\\
Star Field 12 & +12.67 & 116580 & 357 & 18556\\
Star Field 13 & +13.01 & 85468 & 311 & 11625\\
Star Field 14 & +13.32 & 64046 & 305 & 9007\\
Star Field 15 & +14.19 & 28631 & 241 & 2730\\
Star Field 16 & +14.87 & 15402 & 232 & 1751\\
Star Field 17 & +15.89 & 6004 & 251 & 844\\
\hline
\end{tabular}
\end{table*}

\vfill

\end{document}